\documentclass{article}

\usepackage{adjustbox}
\usepackage{float}
\usepackage{graphicx}
\usepackage{multicol}
\usepackage{multirow}
\usepackage[normalem]{ulem}
\usepackage[svgnames]{xcolor}
\usepackage{hyperref}
\usepackage[capitalise]{cleveref}
\usepackage{xurl}
\usepackage{tablefootnote}
\usepackage{pythonhighlight}
\usepackage[frozencache,cachedir=minted-cache]{minted}

\title{Building net-native agreement systems}
\author{Joshua Z. Tan and Luke V. Miller}
\date{May 2022}

\begin{document}

\maketitle

\begin{abstract}
Agreements and contracts are everywhere, but they are built on layers and layers of legal and social institutions. Software is slowly entering into this stack. In this article, we introduce agreement paths, a general model for understanding and decomposing digital agreement systems, and \href{https://github.com/metagov/agreementengine}{Agreement Engine}, an open-source software service for building net-native agreement systems. We demonstrate Agreement Engine by building two example agreement systems: Scarce Knowledge, an app for crowdfunding essays, and Twitter Social Capital, a bot that allows users to form and enforce Twitter agreements.
\end{abstract}

\section{Introduction}

Agreements are the building blocks of modern societies. Whenever you want to work with other people—whether it’s starting a business, building open-source software, or slaying a (virtual) dragon—you will need to first form an agreement with those people to communicate expectations, safeguard rights, or divvy up rewards. When we change the agreements that people can form, for example through policies that affect communication technology, social trust, or the rule of law, we change the kinds of economies and politics that people produce. 

But what determines the agreements that people can form? Or, to put it another way: for a given kind of agreement (legal, informal, computational, etc.), how can we build \textit{systems} for authoring and enforcing such agreements? Such agreement systems range widely: from a state’s business registration process, to the small claims court system, to the Ethereum smart contract system, to the guild systems of online games like World of Warcraft or EVE Online, to informal systems of agreement that govern everything from bazaars to traffic circles.

In this article, we introduce agreement paths, a general formalism for modeling agreement systems, and Agreement Engine, a software service for building net-native agreement systems. In doing so, we emphasize (1) the growing importance of agreement systems that are implemented partially or wholly online and (2) the benefits of a more rigorous and interoperable framework for engineering such systems. We also present two experimental agreement systems, Scarce Knowledge and Twitter Social Capital, built using Agreement Engine. Finally, we contextualize our contributions within the larger project of building net-native agreement systems and close out with a discussion of the user experience of legal contracts compared to smart contracts or informal agreements.

\section{Background and related work}

What do the following have in common?

\begin{enumerate}
	\item A signed rental contract

	\item A one-line email: ``cool, meet you at 9pm outside the bar"

	\item A (multi-party) smart contract

	\item A bid on eBay

	\item A video of two people exchanging nods

\end{enumerate}

Answer: they are all manifestations of mutual assent, a.k.a. \textit{agreements}. \textit{Mutual assent} is just a piece of jargon that means ``we agree"—it’s an abstract relation (what the law calls a ‘meeting of the minds’) between two or more subjects that share a common intention or belief. Agreements are \textit{manifestations} of mutual assent, meaning that they bear witness to the claim that a common intention or belief exists \cite{legal_information_institute_agreement_2021}. To put it another way: every agreement constitutes evidence of mutual assent, though not every such agreement is thereby made enforceable.

Many agreements also incur \textit{obligations}, where an obligation\textit{ }is an action that one has to do. These obligations usually assume certain mechanisms for \textit{enforcement}, where \textit{enforcement} is the act of compelling compliance with an obligation. For example: an email thread between two friends agreeing to meet at a bar (1) is explicit evidence of mutual assent and (2) creates a social obligation to meet at said bar where (3) the obligation to show up is enforced by an implied social norm: don’t break promises to friends or you’ll lose your friends and/or be reputed as a flake. A bid on eBay, on the other hand, (1) is evidence of the bidder’s agreement to buy the item at that price and (2) creates an obligation to pay the seller at that price if they win the auction where (3) that obligation is presumptively legally enforceable (4) either through eBay’s automated escrow system or through eBay’s (less-automated) unpaid item policy. \cite{ebay_inc_ebay_2021}

This (1)-(2)-(3) pattern of agreement-obligation-enforcement is extremely common in society and in the law. We call this pattern a \textit{contract system}, and we call agreements that fit this pattern \textit{contracts}. There is substantial theoretical and empirical work on contract systems within comparative law \cite{reimann_oxford_2006}, contract law \cite{merkin_pooles_2021}, philosophy \cite{markovits_philosophy_2021}, and economics \cite{bolton_contract_2004}, mostly focusing on national legal contract systems. Social scientists have also studied informal agreement systems—informal only in the sense that they do not directly involve a court system—through an institutional lens. \cite{ostrom_governing_1990} And most recently, smart contract systems—systems of agreements automatically enforced through code—have matured from early conceptual models \cite{szabo_smart_1996} into large, blockchain-based platforms such as Ethereum \cite{buterin_next-generation_2013} and Cosmos \cite{kwon_cosmos_nodate}. 

In what follows, we will focus on a number of features shared between traditional legal contract systems, informal agreement systems, and smart contract systems.

\begin{table}[H]
\begin{adjustbox}{max width=\textwidth}
\begin{tabular}{p{4.23cm}p{12.28cm}p{4.23cm}p{12.28cm}}
\hline
\multicolumn{1}{|p{4.23cm}}{\textbf{Term}} & 
\multicolumn{1}{|p{12.28cm}|}{\textbf{Definition}} \\ 
\hline
\multicolumn{1}{|p{4.23cm}}{Agreement} & 
\multicolumn{1}{|p{12.28cm}|}{A manifestation of mutual assent between a number of parties} \\ 
\hline
\multicolumn{1}{|p{4.23cm}}{Obligation} & 
\multicolumn{1}{|p{12.28cm}|}{An action that one has to do} \\ 
\hline
\multicolumn{1}{|p{4.23cm}}{Enforcement} & 
\multicolumn{1}{|p{12.28cm}|}{The act of compelling compliance with an obligation} \\ 
\hline
\multicolumn{1}{|p{4.23cm}}{Contract} & 
\multicolumn{1}{|p{12.28cm}|}{An agreement that creates an obligation with an enforcement mechanism} \\ 
\hline
\multicolumn{1}{|p{4.23cm}}{Contract system} & 
\multicolumn{1}{|p{12.28cm}|}{A system for creating and enforcing contracts} \\ 
\hline
\multicolumn{1}{|p{4.23cm}}{Legal contract} & 
\multicolumn{1}{|p{12.28cm}|}{An agreement recognized and enforced by the state, typically through a system of laws, courts, and enforcement agencies} \\ 
\hline
\multicolumn{1}{|p{4.23cm}}{Smart contract} & 
\multicolumn{1}{|p{12.28cm}|}{An agreement written in code and enforced by a technical system such as a virtual world or blockchain} \\ 
\hline
\multicolumn{1}{|p{4.23cm}}{Informal agreement} & 
\multicolumn{1}{|p{12.28cm}|}{An agreement not intended to be enforced through a legal or technical mechanism\tablefootnote{Informal agreements can include very explicit terms and consequences, and some are actually more diligently enforced than any formal contract, e.g. a loan shark’s ``if you don’t pay me back, I’m going to beat you up". The difference is that there is no formal institutional structure or system for enforcement.} but instead through social norms, reputational incentives, and/or extralegal means that do not depend on a formal institutional enforcement mechanism} \\ 
\hline
\multicolumn{1}{|p{4.23cm}}{Joint account} & 
\multicolumn{1}{|p{12.28cm}|}{A digital representation of an n-person agreement, akin to a user account} \\ 
\hline
\multicolumn{1}{|p{4.23cm}}{Registration} & 
\multicolumn{1}{|p{12.28cm}|}{The process of recording or representing a completed agreement and related data, usually by a qualified authority} \\ 
\hline
\multicolumn{1}{|p{4.23cm}}{Authentication} & 
\multicolumn{1}{|p{12.28cm}|}{The process of checking the data, status, and/or registration of an agreement, usually by an enforcement mechanism} \\ 
\hline
\end{tabular}
\end{adjustbox}
\caption{A glossary of the many terms in this paper.}
\end{table}

\section{Modeling agreement systems}
Imagine that you are driving through the streets of Delhi. A worn-down yellow tuk-tuk beeps; you slow, and it merges in front of you. At the next stop light, you wince as a big garbage truck rumbles to a stop just behind your back window. A young chai wallah weaves through the cars with a tray of plastic cups; you wave him over and trade him a few rupees for a cup. The tuk-tuk ahead of you moves on; you accelerate too, swerving past a bicycle vendor on the corner, and turn onto a crowded traffic circle. As you inch forward, a red hatchback edges up at the next turn, signaling right. You slow, graciously; the hatchback moves gingerly up until it works up the courage and enters ahead of you into the wide stream of cars heading downtown.

Every day, we negotiate agreements with each other. We signal; we gesture; we infer; we agree; we act. It is second nature to us. Most of these rich interactions do not produce well-defined artifacts in the form of enforceable contracts. But most \textit{do} take place within systems that support and facilitate them, systems that have a passing resemblance, if you stare at them closely, to more formal systems like the courts. Our goal in this section is to understand and characterize some of these similarities. 

We present \textit{agreement paths}, a formalism for modeling systems of agreements that abstracts over the content and type of an agreement. While agreement paths are general enough to apply to many types of agreements, they are not intended to be a general analytic instrument for scholars of agreements in philosophy, economics, or law. \textit{Agreement paths are intended to highlight the elements of an agreement system that can be digitized.} Our goal is not to present a general model for analyzing or classifying agreement systems but to model the components that are present in digital agreement systems like DocuSign Agreement Cloud, Ethereum, Aragon Court, or Kleros. This way we can more easily (1) recognize these digital agreement systems when they occur on the internet, (2) amend these systems as we amend institutions for agreement in the real world, (3) identify the places where digital processes can enter into offline agreement systems, and (4) build more modular and ``swappable" agreement systems.

\begin{figure}[H]
\centering
\centerline{\includegraphics[scale=0.75]{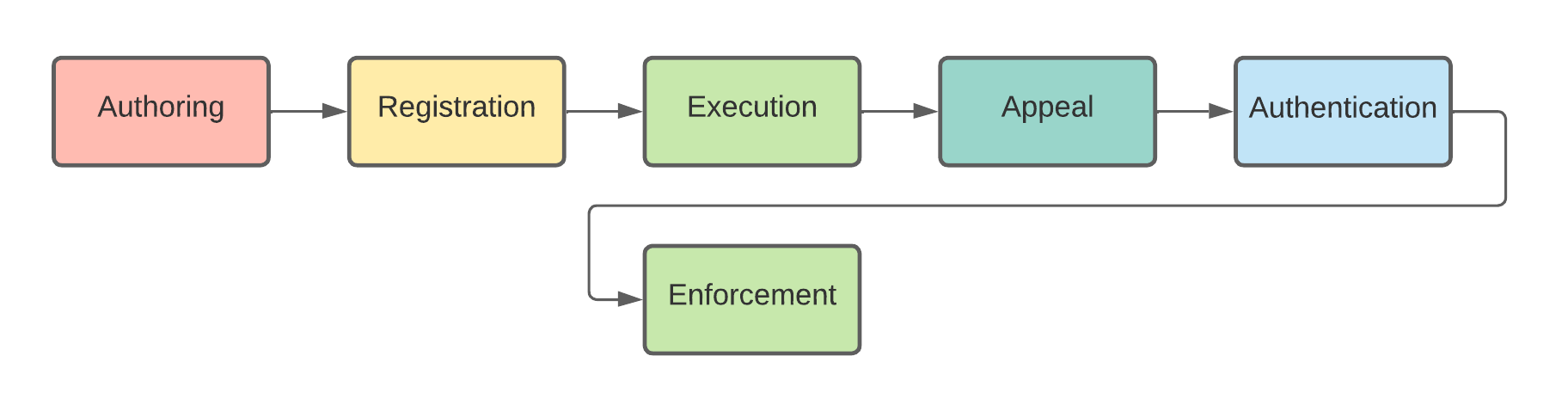}}
\caption{The six stages of an agreement path.}
\label{fig:six_stages}
\end{figure}

Every agreement path can be broken down into six stages: \textit{authoring}, \textit{registration}, \textit{execution}, \textit{authentication}, \textit{appeal}, and \textit{enforcement}. An authoring process allows parties to negotiate and author agreements with each other. A registration process records the data pertaining to an completed agreement, often including the agreement’s identity, content, parties, and status along with secondary authentication data (e.g. the name of a witness). Once registered, the agreement is executed or enacted by the parties to the agreement. During execution, a party may trigger an appeal to exit the execution process for whatever reason (this external appeals process should be distinguished from actions defined within the content of the agreement for resolving disputes).\footnote{ An appeal (or the right to appeal) is a part of the meaning of the agreement that is independent of the ``content" of the agreement. It is enabled by the agreement system.} Following an appeal, an authentication process verifies the data, validity, and/or registration status of an agreement. Finally, an enforcement process interprets the content of an agreement, often but not always with accompanying evidence, in order to enforce its execution or compensate for any outstanding obligations. See \Cref{fig:six_stages}.

For example, Alice and Bob might negotiate the terms of an employment contract over email, trading versions back and forth (authoring). Once they are both satisfied, they convert the final agreement to a PDF and each sign it digitally, trading the signed and counter-signed copies over email (registration). With the contract signed, Bob begins working for Alice, and in return Alice pays Bob a salary, as per the contract (execution). At some point, Alice stops paying Bob and stops responding to calls or email; Bob is forced to file a lawsuit to get back his wages (appeal). As part of this lawsuit, he appends copies of the emails and the signed employment contracts (authentication). The lawsuit enters the legal system. A small claims court rules against Alice, who does not show up to contest the lawsuit, and orders Alice to pay Bob (enforcement) by sending a letter to Alice’s last-known address. Alice does not respond to the letter, at which point the trial court issues a contempt order for Alice to be arrested (appeal).

\begin{figure}[H]
\centerline{\includegraphics[width=14.33cm,height=5.16cm]{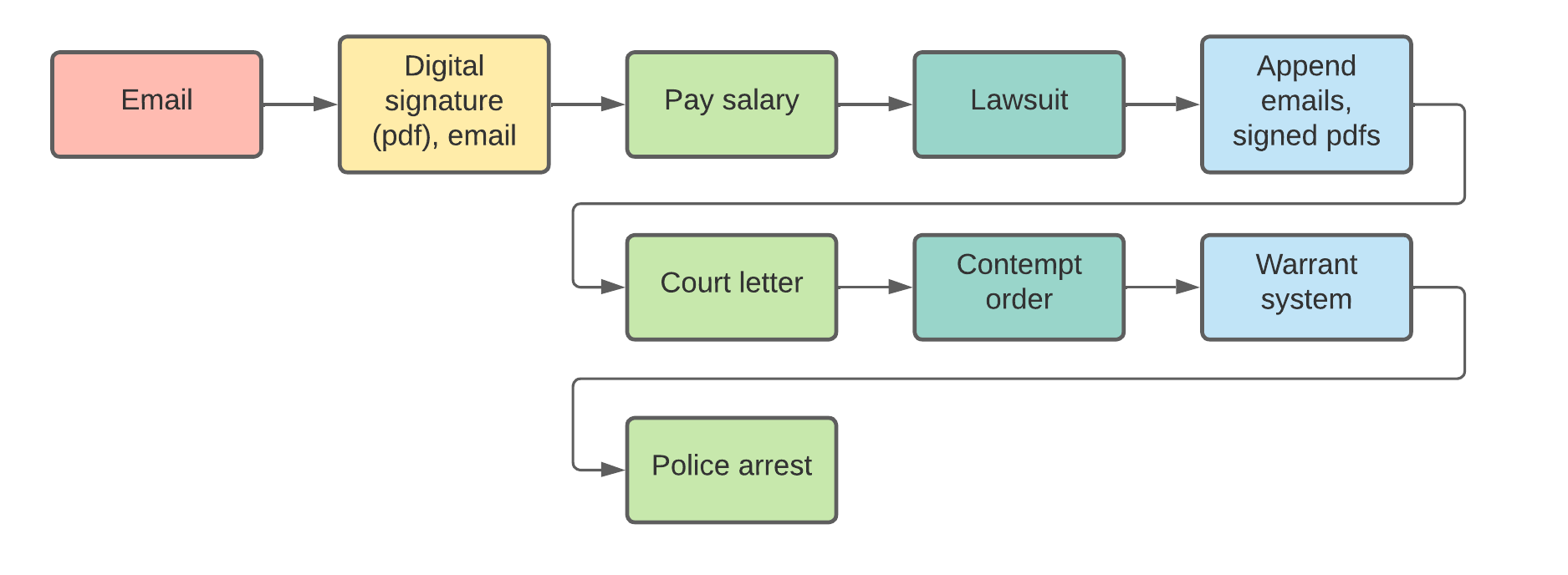}}
\caption{A simple employment contract negotiated over email and enforced via a trial court.}
\end{figure}

Not all agreement paths are as straightforward as the one above.

For example, Ghost Knowledge \cite{azout_ghost_nodate} is a website that allows people to crowdfund ``bounties" for specific authors or researchers to write essays. Once pledges go past a certain threshold, Ghost Knowledge messages the authors on social media, who then either accept or reject the bounty; if they accept, the money is put into an escrow account to be released once the essay is published.

\begin{figure}[H]
\centering
\centerline{\includegraphics[width=11.46cm,height=14.61cm]{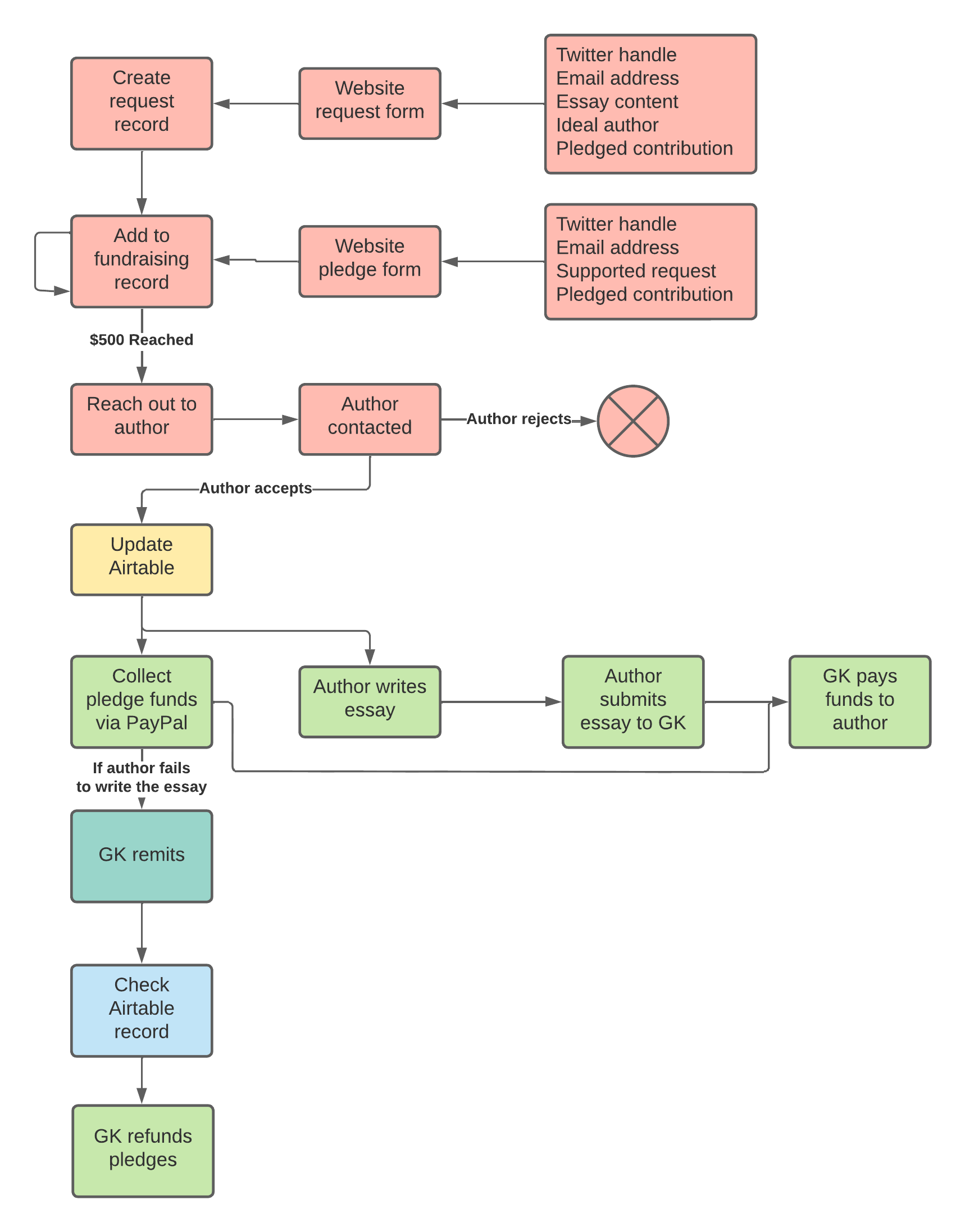}}
\caption{The agreement path of Ghost Knowledge, a website for crowdfunding written essays.}
\end{figure}

Ghost Knowledge was built and released using no-code tools including Airtable, PayPal, and the Twitter API. A substantial portion of the author interactions involved manual messaging by the founders. \cite{azout_interview_2021} We will come back to Ghost Knowledge in the next section, when we recreate a version of it using the Agreement Engine tool.

\begin{table}[H]
\begin{adjustbox}{max width=\textwidth}
\begin{tabular}{p{3.15cm}p{7.75cm}p{5.42cm}p{3.15cm}p{7.75cm}p{5.42cm}}
\hline
\multicolumn{1}{|p{3.15cm}}{\textbf{Scope}} & 
\multicolumn{1}{|p{7.75cm}}{\textbf{Definition}} & 
\multicolumn{1}{|p{5.42cm}|}{\textbf{Examples}} \\ 
\hline
\multicolumn{1}{|p{3.15cm}}{Authoring} & 
\multicolumn{1}{|p{7.75cm}}{An authoring process allows end-users to negotiate and author agreements with each other. } & 
\multicolumn{1}{|p{5.42cm}|}{Ghost knowledge negotiation (No-code form + Twitter DMs)} \\ 
\hline
\multicolumn{1}{|p{3.15cm}}{Registration} & 
\multicolumn{1}{|p{7.75cm}}{The process of recording or representing a completed agreement and related data, usually by a qualified authority} & 
\multicolumn{1}{|p{5.42cm}|}{Signing a physical document, clicking accept on a digital form} \\ 
\hline
\multicolumn{1}{|p{3.15cm}}{Execution} & 
\multicolumn{1}{|p{7.75cm}}{A special layer of the enforcement process related to the content of an agreement. Includes things such as enactment, delivery, and any number of triggers specified in the content of the agreement, e.g. revisions, termination clauses, and so forth.} & 
\multicolumn{1}{|p{5.42cm}|}{Payment, sending of goods, execution of contract calls} \\ 
\hline
\multicolumn{1}{|p{3.15cm}}{Appeal} & 
\multicolumn{1}{|p{7.75cm}}{A process that triggers an authentication and enforcement process. Appeals can be pre-specified within an agreement, but they always call an external process that is not specified in the agreement itself.} & 
\multicolumn{1}{|p{5.42cm}|}{Communication between parties to revise a contract. \newline
\textit{Non-example: an adjudication clause specified in the contract itself.}} \\ 
\hline
\multicolumn{1}{|p{3.15cm}}{Authentication} & 
\multicolumn{1}{|p{7.75cm}}{The process of checking the data, status, and/or registration of an agreement, usually by an enforcement mechanism} & 
\multicolumn{1}{|p{5.42cm}|}{Admission of evidence in a courtroom, reporting a video on YouTube} \\ 
\hline
\multicolumn{1}{|p{3.15cm}}{Enforcement} & 
\multicolumn{1}{|p{7.75cm}}{An enforcement mechanism interprets the content of an authenticated agreement, often (but not always) with accompanying evidence, in order to enforce or compensate for any outstanding obligations.} & 
\multicolumn{1}{|p{5.42cm}|}{Kleros jury, U.S. justice system, moderator intervention on a social media platform} \\ 
\hline
\end{tabular}
\end{adjustbox}
\caption{Definitions of the stages of an agreement path.}
\end{table}

\begin{figure}[H]
\centering
\centerline{\includegraphics[width=14.33cm,height=7.43cm]{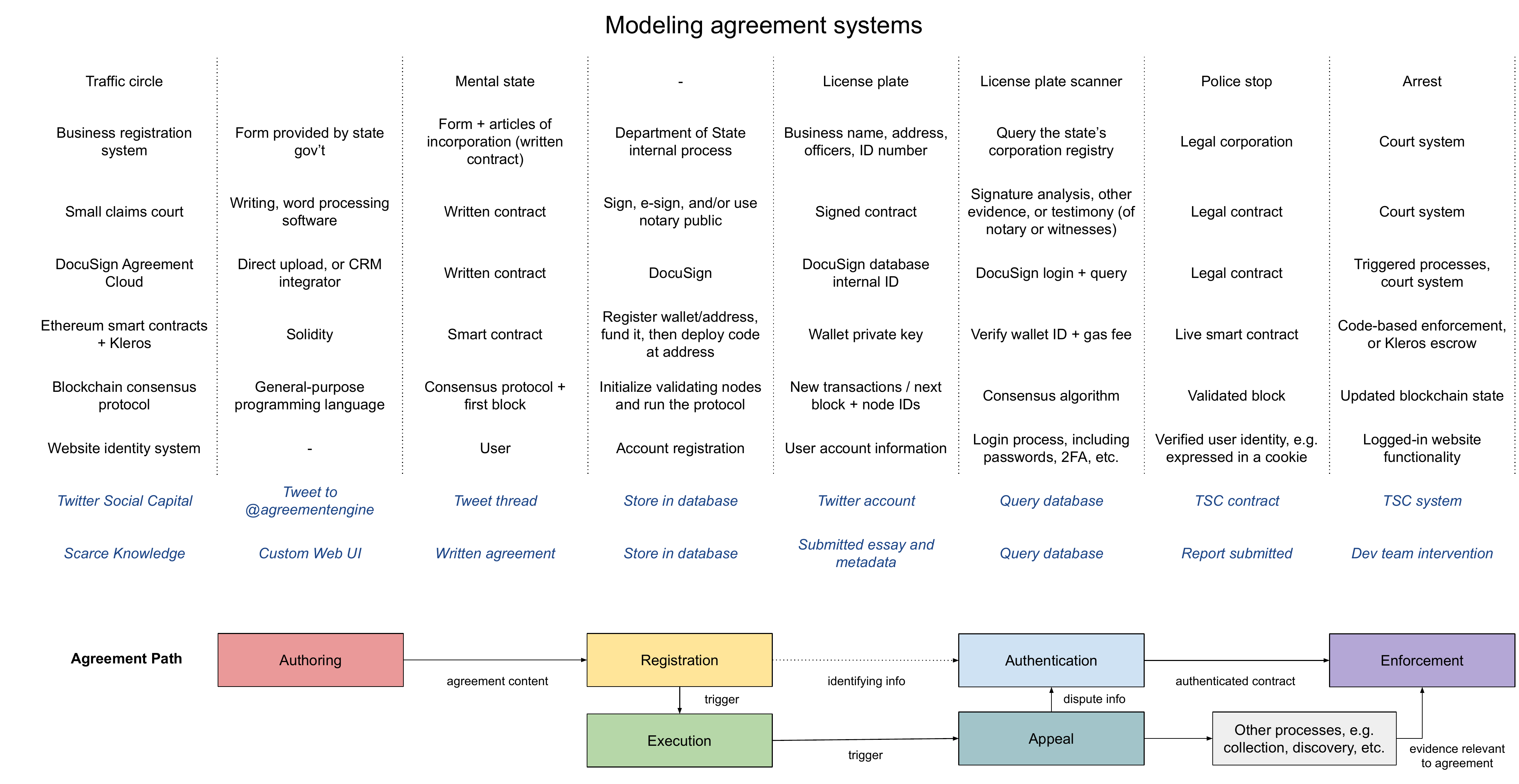}}
\caption{Additional examples of the key processes in an agreement system: \textit{authoring}, \textit{registration}, \textit{execution}, \textit{appeal, authentication}, and \textit{enforcement}.}
\end{figure}

\section{Constructing and composing agreement paths}
Agreement paths, on their own, are a conceptual model intended to support the composition and substitution of different processes within agreement systems. In this section, we present \href{https://github.com/metagov/agreementengine}{Agreement Engine}, a software tool for building net-native agreement systems. Agreement Engine implements the agreement path model.

To be clear, Agreement Engine itself is neither a system for authoring individual agreements nor a system for enforcing them. However, it has built-in interfaces with existing authoring and enforcement systems.

\subsection{Agreement processes}
From a top-down perspective, each agreement system created in Agreement Engine can be considered a state machine composed of many processes. Each process represents a distinct and independent service: authoring, registration, authentication, and so on. Within a given path, not all processes need to be computational, as long as they have a formal representation. When a process is active, all data input to the agreement is routed to it for processing. The process can then decide to wait for additional input, transition to another process, or terminate the agreement. Processes can also take action immediately when they are activated or right before the process ends, instead of asynchronously upon receiving input. Using this method, we can chain discrete processes into agreement paths\textbf{\textit{ }}which exhibit more complex behavior. 

In our archetypical agreement system model, this starts with an \textit{Authoring} process and ends with an \textit{Enforcement }process\textit{.}

\begin{figure}[H]
\centering
\includegraphics[width=9.86cm,height=8.61cm]{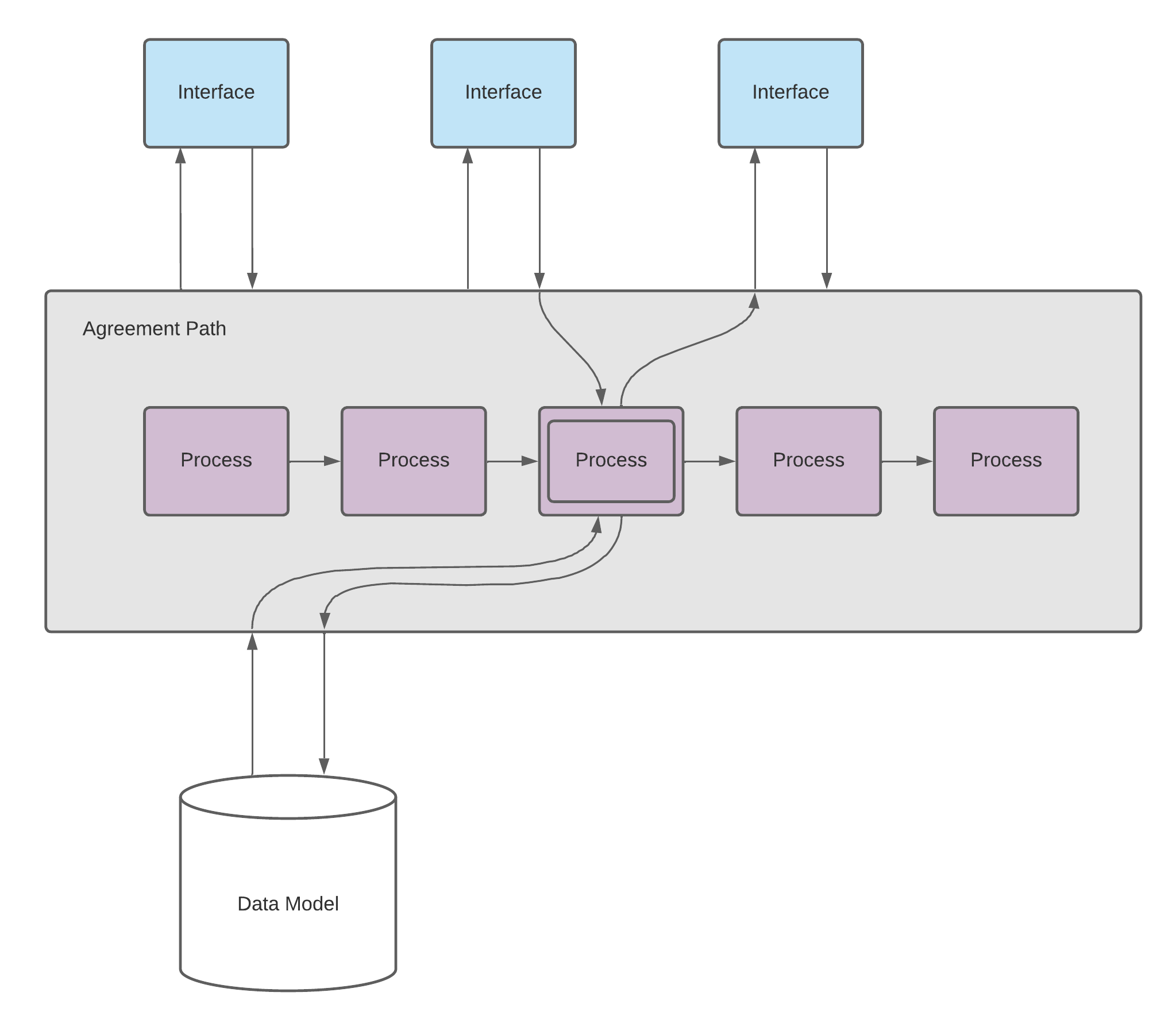}
\caption{Diagram of a single agreement instance and connected systems}
\label{fig:single_agreement_instance}
\end{figure}

\subsection{Agreement interfaces and instances}

\begin{figure}[H]
\centering
\centerline{\includegraphics[width=14.33cm,height=7.64cm]{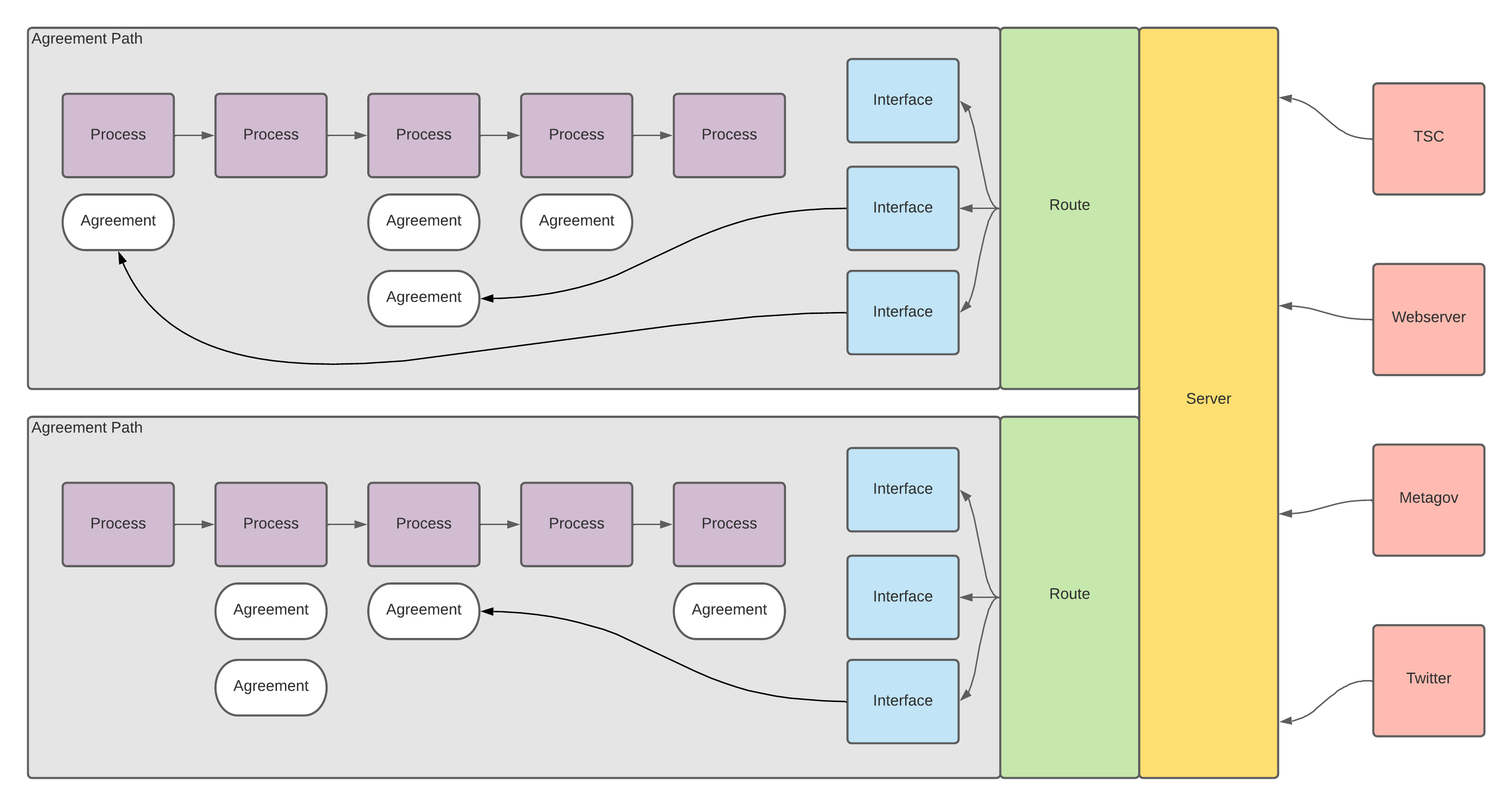}}
\caption{Diagram of a server with multiple paths and active agreements}
\label{fig:server_multiple_paths}
\end{figure}

Within Agreement Engine, agreement paths are high-level data structures that model a single agreement system along with an arbitrary number of agreements authored within that system. Paths can manage processes by keeping track of which process is currently active, handling state transitions, and providing access to each agreement’s private data model. Agreement paths also filter and redirect agreement input and output via a set of agreement interfaces. Interfaces filter incoming data from various sources so that it can be preprocessed and routed to the correct agreement instance (or used to create a new agreement instance). As seen in \Cref{fig:single_agreement_instance}, a single instance of an agreement path, or simply referred to as an agreement instance, acts as a passthrough for the currently-active process. It allows the agreement to be viewed as a single object from the outside, while facilitating more complex decision-making from the inside. Each agreement instance has its own state and data model in order to track the path it is associated with,  the stage within that path, and any additional data relevant to its execution. Paths can manage any number of agreements, as seen in \Cref{fig:server_multiple_paths}.

\subsection{Agreement servers}

Agreement servers allow for multiple paths (and thus multiple agreement instances) to be run at the same time. The server receives requests and routes them to the corresponding path where the interfaces decide which agreement instance should receive the data. This system allows for arbitrary data to be handled, whether it's from a local process, a webhook, or another web server. The server provides the infrastructure for agreements to connect to the internet and interface with other platforms and software. For example, suppose we have a server running on a web server with the domain name ``\textit{myagreements.com}." We could configure a new agreement path called ``\textit{Employment}" for managing employment contracts. Now our agreements can receive data sent to ``\textit{myagreements.com/Employment}" via REST API calls from other applications, or webhooks from platforms such as Twitter or Slack.

\subsection{Example: Scarce Knowledge}

\begin{figure}[H]
\centering
\includegraphics[width=9.72cm,height=11.57cm]{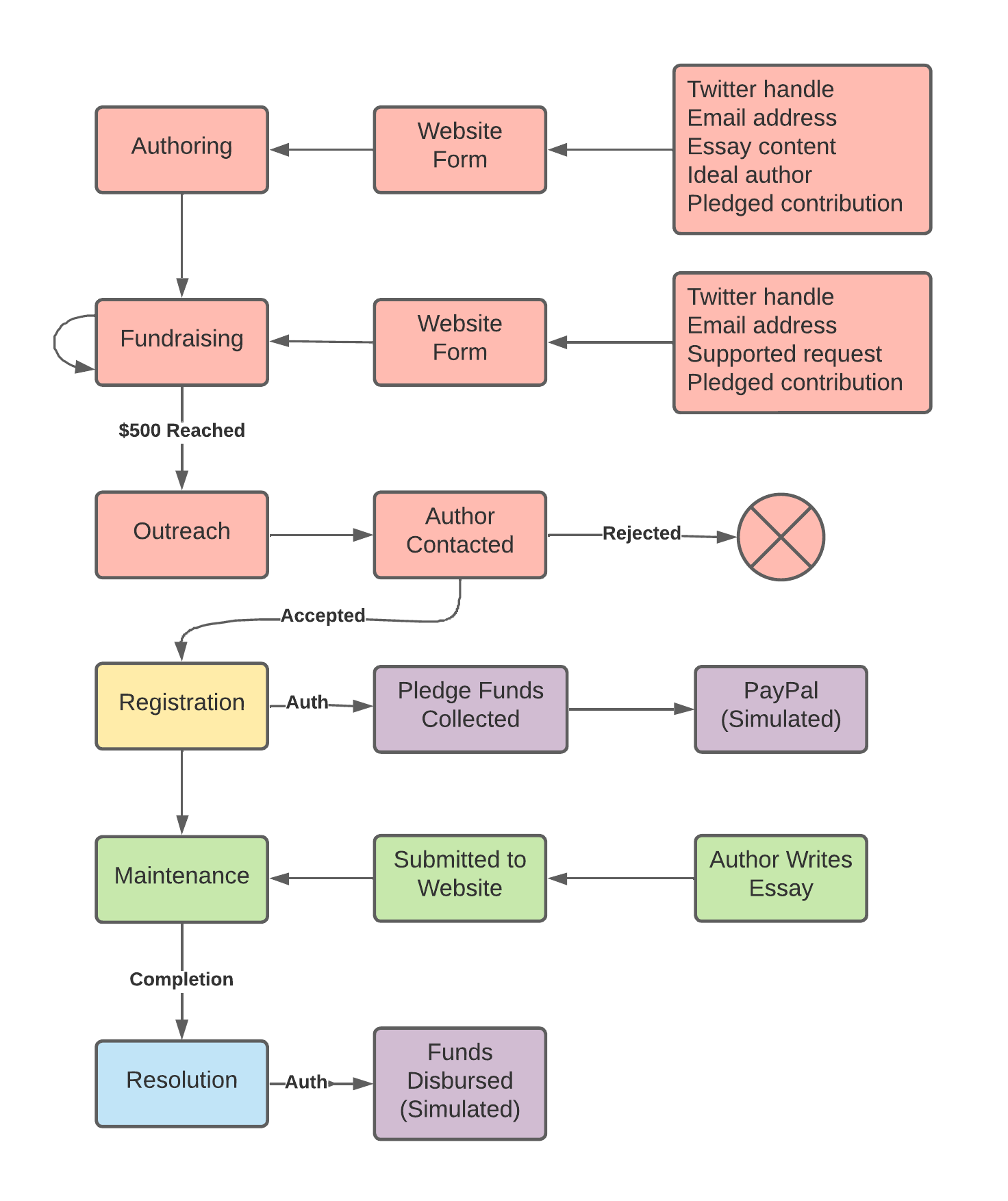}
\caption{The agreement path for Scarce Knowledge}
\label{fig:scarce_knowledge_path}
\end{figure}

Let’s look at an example agreement system built using Agreement Engine. This demo, which we call Scarce Knowledge, recreates a version of the \href{https://www.ghostknowledge.com/}{Ghost Knowledge} platform that we described in the previous section. \Cref{fig:scarce_knowledge_path} shows the agreement path of this version. Requests to Scarce Knowledge are initiated by a user-facing website with a simple form that forwards data to Agreement Engine. Once pledged contributions surpass $\$$500, a process transition is triggered. This is done via a Twitter interface that sends a direct message to the requested author. The next few processes return to the website to handle final essay submission. Finally, an email automation service is called to forward the essay to those who supported it. By using Agreement Engine, we were able to manage a complex decision-making process with several approval steps that span a user facing frontend, Twitter, and a mail automation service. In a production implementation this would also include PayPal integration. \Cref{fig:scarce_knowledge_interface} below shows part of the user interface flow for this agreement path implementation using a simple website and Twitter direct messages.

\begin{figure}[htbp]
\centerline{\includegraphics{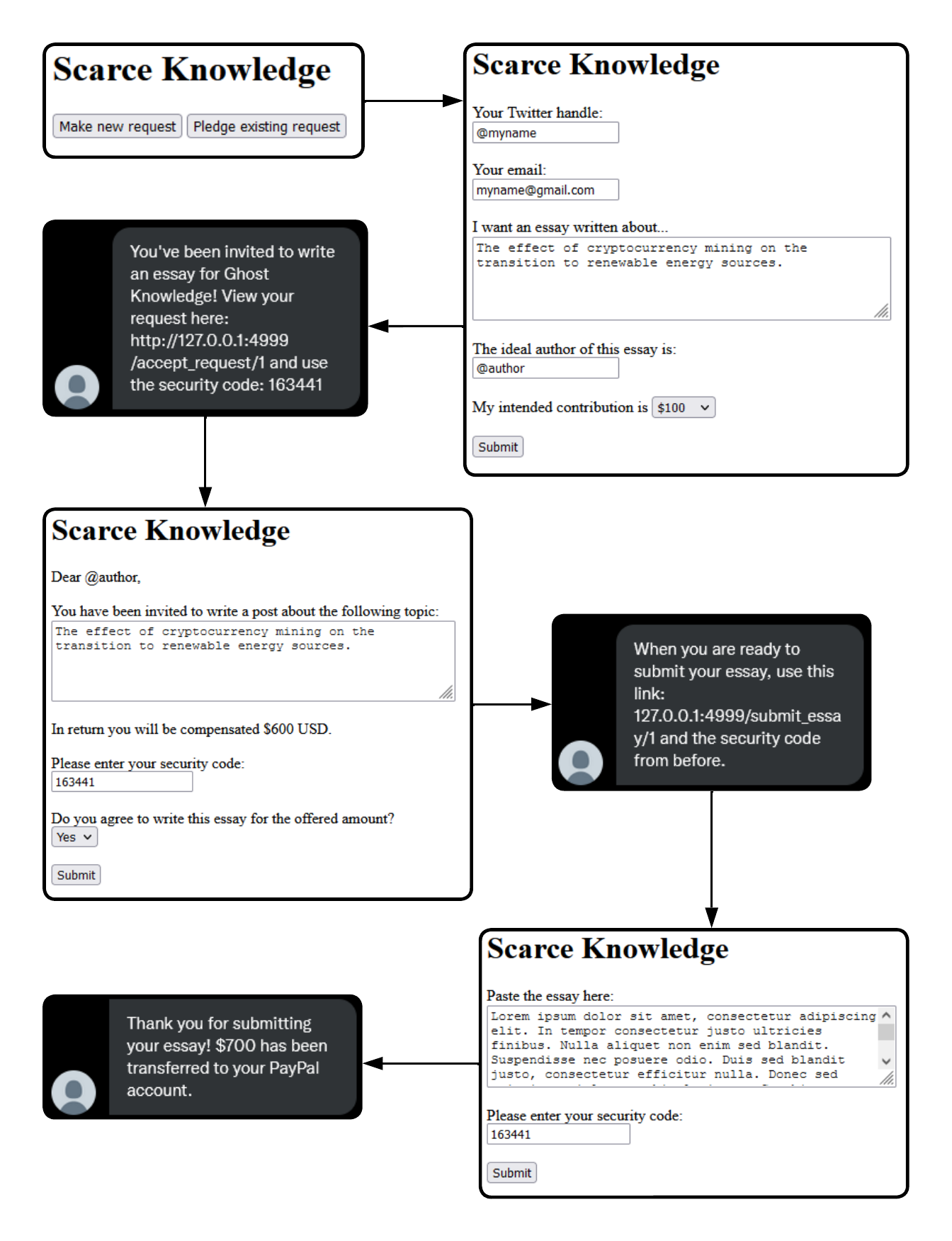}}
\caption{User interface diagram for ``Scarce Knowledge"}
\label{fig:scarce_knowledge_interface}
\end{figure}

\subsection{Example: Twitter Social Capital}

\begin{figure}[htbp]
\centering
\includegraphics[width=10.97cm,height=13.45cm]{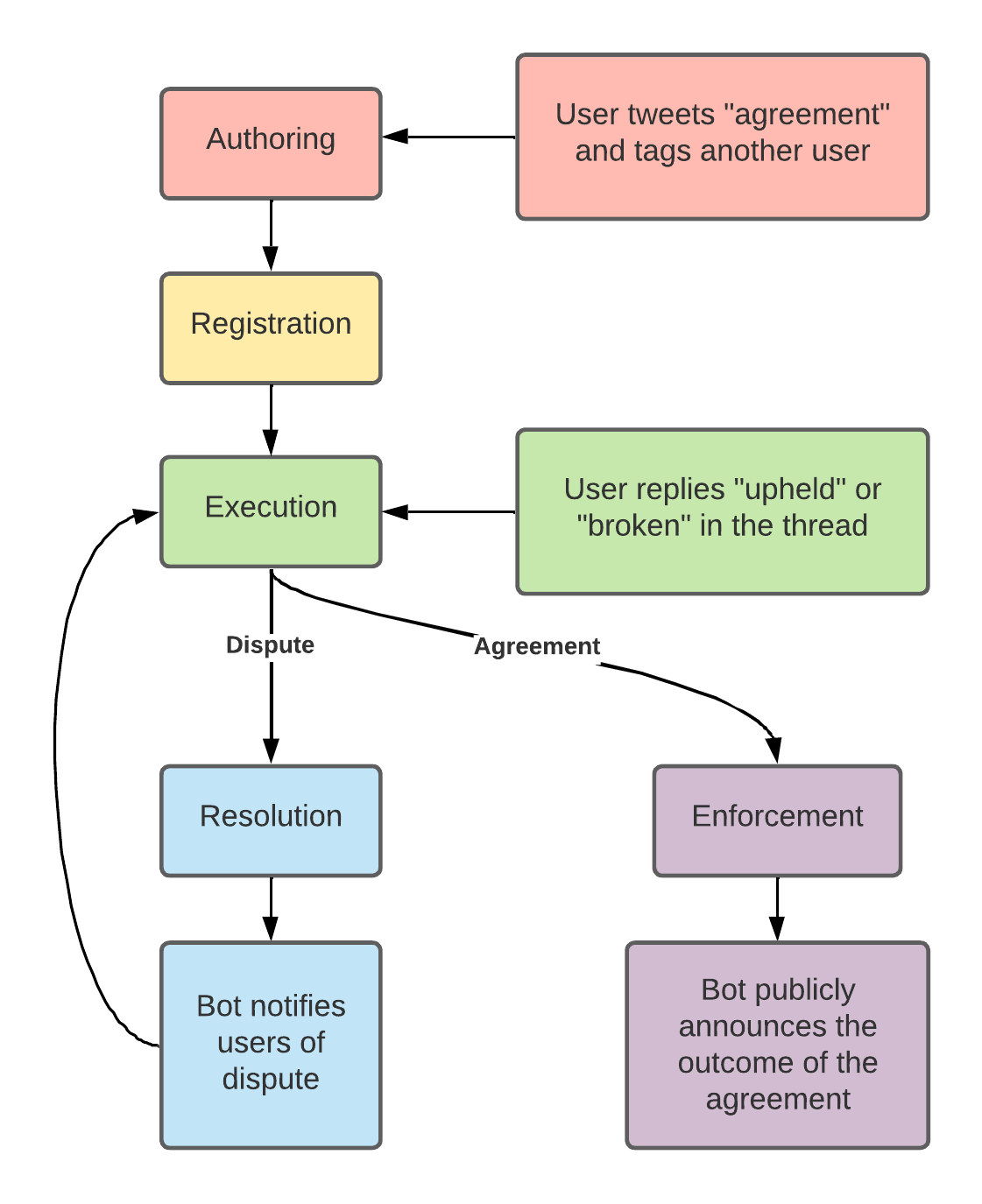}
\caption{The agreement path for Twitter Social Capital (TSC)}
\label{fig:tsc_agreement_path}
\end{figure}

Let’s look at another example agreement system implemented in Agreement Engine. \href{https://medium.com/metagov/introducing-the-agreement-engine-bf03b6d5c16c}{Twitter Social Capital} (TSC) is a Twitter-based agreement system that allows users to register and enforce agreements with each other by tweeting at the \href{https://twitter.com/agreementengine}{@agreementengine} bot. New agreements are authored by tagging another user and using the keyword ``agreement" Users can then reply to the tweet with ``upheld" or ``broken" to indicate the status of the agreement. If users disagree on the outcome they are given the chance to change their responses and come to a consensus. Finally, the bot publicly records the result of the agreement. As seen in \Cref{fig:tsc_agreement_path}, by using Agreement Engine we were able to create a branching and looping control flow to resolve disputes and only terminate when an agreement is reached. This example uses an interface to the \href{https://gateway.metagov.org/}{Metagov Gateway} to send and receive tweets. The technical details of interfaces are discussed later, but this implementation demonstrates the flexibility of Agreement Engine in integrating with other platforms.

\begin{figure}[H]
\centerline{\includegraphics[width=14.33cm,height=13.29cm]{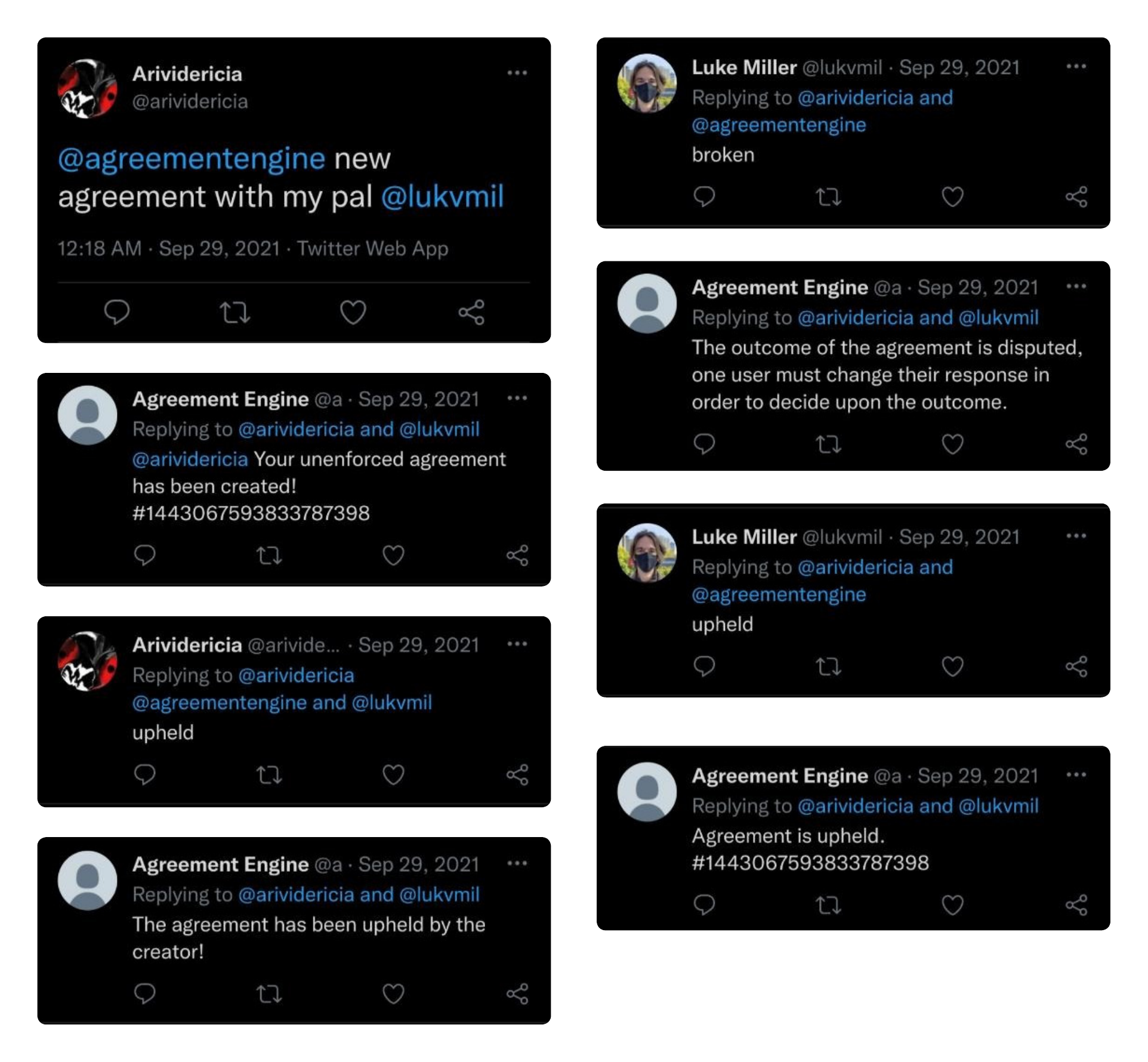}}
\caption{Example tweets from the TSC Twitter bot that we built using the Agreement Engine.}
\end{figure}

\begin{table}[H]
\begin{adjustbox}{max width=\textwidth}
\begin{tabular}{p{4.13cm}p{4.13cm}p{4.13cm}p{4.13cm}p{4.13cm}p{4.13cm}p{4.13cm}p{4.13cm}}
\hline
\multicolumn{1}{|p{4.13cm}}{\textbf{Feature}} & 
\multicolumn{1}{|p{4.13cm}}{\textbf{Notary public}} & 
\multicolumn{1}{|p{4.13cm}}{\textbf{DocuSign Agreement Cloud}} & 
\multicolumn{1}{|p{4.13cm}|}{\textbf{Kleros or Aragon Agreements}} \\ 
\hline
\multicolumn{1}{|p{4.13cm}}{Supports authoring} & 
\multicolumn{1}{|p{4.13cm}}{No} & 
\multicolumn{1}{|p{4.13cm}}{Yes, through Gen and other integrators} & 
\multicolumn{1}{|p{4.13cm}|}{No} \\ 
\hline
\multicolumn{1}{|p{4.13cm}}{For legal consumption} & 
\multicolumn{1}{|p{4.13cm}}{Yes} & 
\multicolumn{1}{|p{4.13cm}}{Yes} & 
\multicolumn{1}{|p{4.13cm}|}{No} \\ 
\hline
\multicolumn{1}{|p{4.13cm}}{For consumption by online community} & 
\multicolumn{1}{|p{4.13cm}}{No} & 
\multicolumn{1}{|p{4.13cm}}{No} & 
\multicolumn{1}{|p{4.13cm}|}{Yes} \\ 
\hline
\multicolumn{1}{|p{4.13cm}}{Many enforcement mechanisms} & 
\multicolumn{1}{|p{4.13cm}}{No, just the law} & 
\multicolumn{1}{|p{4.13cm}}{No, just the law} & 
\multicolumn{1}{|p{4.13cm}|}{No, just escrow} \\ 
\hline
\multicolumn{1}{|p{4.13cm}}{Many authoring platforms} & 
\multicolumn{1}{|p{4.13cm}}{No} & 
\multicolumn{1}{|p{4.13cm}}{Some, e.g. Salesforce} & 
\multicolumn{1}{|p{4.13cm}|}{No, just Ethereum} \\ 
\hline
\multicolumn{1}{|p{4.13cm}}{Includes decision-making process for enforcement} & 
\multicolumn{1}{|p{4.13cm}}{No} & 
\multicolumn{1}{|p{4.13cm}}{No} & 
\multicolumn{1}{|p{4.13cm}|}{Yes, digital jury} \\ 
\hline
\end{tabular}
\end{adjustbox}
\caption{Comparison of different registration services.}
\end{table}

\section{Using the Agreement Engine library}

When creating an agreement system using the \href{https://github.com/metagov/agreementengine}{Agreement Engine Python library}, there are three main components that should be considered: the server, paths, and interfaces. Let’s construct a simple agreement to explore how this works.

\begin{figure}[H]
\begin{minted}[frame=lines]{python}
class MyInterface(Interface):
    def filter(self):
        return True
        
    def match(self):
        return self.new_agreement()
\end{minted}
\caption{Example interface code}
\end{figure}

First we need to define an interface. This is how a path decides whether incoming data should be routed to an existing agreement or a new one should be created. In the example code above, we create an interface that handles all input, and always creates a new agreement. Real applications will decide when different interfaces should govern input data, and how to match existing agreements. For example, the Scarce Knowledge example uses two interfaces to communicate with the user facing web server and the Twitter API.

\begin{figure}[H]
\begin{minted}[frame=lines]{python}
class MyPath(AgreementPath):
    class Authoring(AgreementProcess):
        def on_receive(self, data):
            self.model.set('data', 'received', data)
            self.path.terminate()
 
    interfaces = [
        MyInterface
    ]
 
    init = Authoring
\end{minted}
\caption{Example agreement path code}
\end{figure}

Next we define our agreement path, this is the primary structure that represents our agreement system. It will contain all of our processes and define the path that an agreement takes through them. In this example we have a single authoring process that records the data we received to the agreement’s data model (a simple JSON format with different fields) and then terminates. We also set up the interfaces that the path is using, and the initial process that a new agreement will start in. Paths can implement more complex logical flow such as branches or loops. Our example paths use four processes that can terminate early upon detecting invalid data inputs.

\begin{figure}[htbp]
\begin{minted}[frame=lines]{python}
server = Server('db.json')
server.add_path(MyPath)
server.run()
\end{minted}
\caption{Example server code}
\end{figure}

\vspace{1\baselineskip}
Finally we create the server that actually receives and sends requests. All we need to do is define a JSON file to store our agreement data and add our paths. Servers can support any number of paths, as long as they have unique names. Each path will receive a unique API URL for receiving requests. The URL for a path on a local server will look like http://\textit{127.0.0.1/MyPath}.

\section{The challenge of composition and re-use}

One of the first lessons we learned was that enforcement mechanisms, or any agreement process, are not as hot-swappable as we would like to think that they are. The Agreement Engine Python library supports reusing processes by ``transplanting" them from existing agreement paths into new ones, allowing us to build up a library of quickly usable processes, but there are some limitations. When modeling these systems conceptually, we can abstract away from the exact data being passed around by different functions, but in order to implement the system in software, we need to explicitly state what data is being stored in an agreement, accessed in different processes, or passed to external functions. But this data resists standardization. 

To see the problem, imagine a simple example where we want to swap the authoring processes of Scarce Knowledge and TSC. Scarce Knowledge receives new agreements via form submission on a website, while TSC receives new agreements by tweeting at the @agreementengine account. These two methods are both simple and have the same output types (agreements), but problems emerge if we attempt to swap them. The TSC agreement system works by reading tweets that reply to a root tweet, using it to identify unique agreement interactions, and all interactions remain on Twitter. On the other hand, Scarce Knowledge ties together a few different platforms: a simple website, Twitter direct messages, and an email plugin, and the actual content of the agreement has to be synthesized from these platforms. In other words, it is hard to design processes to be completely disentangled from the ones that come before or after them. 

Here we come to one of the fundamental tradeoffs: immediate flexibility vs long-term interoperability. Currently, Agreement Engine provides useful software abstractions including the interface and path model which make it easy to set up simple agreement systems. Users don’t have to worry about the details of how the web server functions, or how to get data to specific agreements. Instead they can focus on the actual logic of the agreement, and what decisions should be made based on the incoming data. In future versions of Agreement Engine, we want to introduce similar software abstractions to path composition—i.e. other words, we want to standardize the interfaces and possible interactions between agreement stages. From a systems perspective, we believe that the benefit of interoperability between different agreement paths is more important than preserving the short-term flexibility of being able to code whatever logic one likes.

\section{Discussion}

The first version of Agreement Engine was a net-native service for \textit{registering} net-native agreements, especially on Twitter, and in a way it still serves that purpose.\footnote{ There is a subtlety here. Within any agreement path there will be a registration process that tracks whether something is an agreement. This registration may store the tracking data anywhere, including outside of the Agreement Engine. However, for practical purposes an agreement path also needs to represent and track the state of an agreement as it passes between stages. So, in effect, Agreement Engine functions as a registration service.} It only evolved into a tool for implementing agreement \textit{systems} as we tried to navigate the difficulties of creating, testing, and swapping net-native enforcement mechanisms for these agreements.

We discussed the swapping problem above. But engaging with agreement systems has also led us to a range of more social and less technical questions. Would different systems for agreement foster new solutions to persistent coordination and collective action problems? How do we reason about the tradeoffs between these systems? And by better understanding the properties and affordances of such systems, could we generate new kinds of agreements with novel properties and affordances? And on the practical side: who will be the first users of something like Agreement Engine?\footnote{ Our hypothesis is that our first users will be communities who want to build customized contract frameworks such as Ghost Knowledge that help produce highly-specific contracts for a particular community.}

Our approach to engaging with these questions is empirical—through more experiments and better experiments. After all, Agreement Engine is ultimately a tool for generating and experimenting with new agreement systems. Here are some other agreement systems that we would (eventually) like to build with it:

\begin{enumerate}
	\item A Twitter agreement, enforced by staking one’s reputation on Reddit

	\item A Markdown file with a code of conduct hosted on GitHub, enforced by a GitHub Group’s owner

	\item A text contract published via Google Doc, enforced by a jury of Slack users

	\item A smart contract \textit{not} on Ethereum, enforced by placing tokens in escrow on Kleros

	\item A grant agreement from one DAO to another DAO, enforced by a third DAO

\end{enumerate}

\section{Acknowledgements}

We would like to thank Lawrence Lessig, Nathan Schneider, and Miriam Ashton for helpful comments in the development of this project.

\section{Bibliography}
\bibliographystyle{unsrt}
\bibliography{references}

\begin{thebibliography}{10}

\bibitem{legal_information_institute_agreement_2021}
Legal~Information Institute.
\newblock Agreement, December 2021.

\bibitem{ebay_inc_ebay_2021}
eBay Inc.
\newblock {eBay} unpaid item policy, December 2021.

\bibitem{reimann_oxford_2006}
Mathias Reimann and Reinhard Zimmerman.
\newblock {\em The {Oxford} {Handbook} of {Comparative} {Law}}.
\newblock Oxford University Press, November 2006.
\newblock Publication Title: The Oxford Handbook of Comparative Law.

\bibitem{merkin_pooles_2021}
Robert Merkin and Séverine Saintier.
\newblock {\em Poole's {Textbook} on {Contract} {Law}}.
\newblock Oxford University Press, Oxford, New York, 15th edition, August 2021.

\bibitem{markovits_philosophy_2021}
Daniel Markovits and Emad Atiq.
\newblock Philosophy of {Contract} {Law}.
\newblock In Edward~N. Zalta, editor, {\em The {Stanford} {Encyclopedia} of
  {Philosophy}}. Metaphysics Research Lab, Stanford University, winter 2021
  edition, 2021.

\bibitem{bolton_contract_2004}
Patrick Bolton and Mathias Dewatripont.
\newblock {\em Contract {Theory}}.
\newblock MIT Press, Cambridge, MA, USA, December 2004.

\bibitem{ostrom_governing_1990}
Elinor Ostrom.
\newblock {\em Governing the {Commons}: {The} {Evolution} of {Institutions} for
  {Collective} {Action}}.
\newblock Cambridge University Press, 1st edition edition, November 1990.

\bibitem{szabo_smart_1996}
Nick Szabo.
\newblock Smart contracts: building blocks for digital markets.
\newblock {\em EXTROPY: The Journal of Transhumanist Thought}, 18(2), 1996.

\bibitem{buterin_next-generation_2013}
Vitalik Buterin.
\newblock A next-generation smart contract and decentralized application
  platform, 2013.

\bibitem{kwon_cosmos_nodate}
Jae Kwon and Ethan Buchman.
\newblock Cosmos {Whitepaper}.

\bibitem{azout_ghost_nodate}
Sari Azout.
\newblock Ghost {Knowledge}.
\newblock Accessible at
  https://web.archive.org/web/20210708204302/https://www.ghostknowledge.com/.

\bibitem{azout_interview_2021}
Sari Azout.
\newblock Interview with {Sari} {Azout}, September 2021.

\end{thebibliography}

\end{document}